\documentstyle[aps,pre]{revtex}
\newcommand{\be}{\begin{equation}}
\newcommand{\ee}{\end{equation}}
\newcommand{\np}{N_{_{\rm PEO}}}

\newcommand{\npt}{N_{_{\rm 2 }}}
\newcommand{\nps}{N_{_{\rm PS}}}
\newcommand{\Sps}{\Sigma_{_{\rm PS}}}
\newcommand{\phips}{\phi_{_{\rm PS}}}
\newcommand{\lps}{l_{_{\rm S}}}
\newcommand{\lpeo}{l_{_{\rm EO}}}
\newcommand{\dz}{\,{\rm d}z}
\newcommand{\pdf}{{\em pdf}}
\begin{document}
\bibliographystyle{prsty1}
\title{Monolayers of Diblock Copolymer at the
Air-Water Interface: \\
The Attractive Monomer-Surface Case}
 \author{M. C. Faur\'e$^{(1)}$, P. Bassereau$^{(1)}$, M.
A. Carignano$^{(2)}$, 
I. Szleifer$^{(3)}$, Y. Gallot$^{(4)}$ 
and D. Andelman$^{(5)}$ }
 \address{$^{(1)}$Physico-chimie Curie, UMR 168 CNRS,
Institut Curie \\ 
11, rue Pierre et Marie Curie, 75231 Paris Cedex 05 France\\
$^{(2)}$Department of Chemical Engineering, University of Delaware, 
Newark, DE 
19716, USA\\
$^{(3)}$Department of Chemistry, Purdue University, West Lafayette, IN 
47907-1393, USA\\
$^{(4)}$Institut C. Sadron, 6, rue Boussingault, 67083 Strasbourg cedex, 
France\\
$^{(5)}$School of Physics and Astronomy, Raymond and Beverly Sackler
Faculty of Exact Sciences\\
 Tel Aviv University, Ramat Aviv 69978 Israel}
 
\date{June 4, 1997}
\maketitle
 
\begin{abstract}
 
We have studied both experimentally and 
theoretically the surface pressure isotherms of copolymers 
of polystyrene-polyethyleneoxide 
(PS-PEO) at the air-water interface. The SCMF (single chain 
mean-field) theory provides a 
very good agreement with the experiments for the entire range
of surface densities 
and shows that the adsorption 
energy per PEO monomer at the air-water interface is about one $k_B T$. 
In addition,
the chain density profile has been calculated for
a variety of surface densities, from the dilute to the 
very dense ones.
The SCMF approach
 has been complemented by a mean-field approach in the low density 
regime, where the PEO chains act as a two-dimensional layer.
 Both theoretical calculations agree with the experiments in 
this region.
 
\end{abstract}
 

\section{Introduction}
 
Insoluble monomolecular layers at interfaces 
have been the
subject of many experimental and theoretical studies since 
the pioneering works of I. Langmuir\cite{gaines,adamson,gelbart}. 
These studies
have been performed on 
small amphiphilic molecules such as surfactants and lipids 
at the gas-liquid or  solid-liquid
interface.

More recently, 
the structure of quasi two-dimensional 
layers of
long polymer chains at interfaces 
have been investigated theoretically 
and  experimentally. 
The polymer layers can be obtained in three different
situations: (i) the polymer chains are grafted to the interface
by one end. This scenario
has inspired many theoretical studies
\cite{mil.sci.251.905,hal.aps.100.31,hal.blobo,gre.binder,szl.review}
especially in the
high surface density regime where elongated 
chain conformations called 
{\it ``brushes''}
are expected. Experimental studies
\cite{ada.jcp.105.4798,aur.prl.66.719,aur.mac.24.2523}
exist in the brush regime as well for the 
low and intermediate 
surface grafting densities.

(ii) In another scenario
the polymer chains 
are adsorbed onto the interface from a liquid solution 
due to weak attractive interactions with the interface.
Concentration profiles of the polymers as a function
of the distance from the interface have been
measured using neutron scattering and neutron reflectivity 
for polymers adsorbed from solution onto 
a solid substrate
\cite{sta.csa.86.143,fie.mac.25.434,sat.mac.23.3860,ami.mac.28.3125}
and
onto an air-liquid interface\cite{gui.jcp.95.4632}.
This scenario was also addressed theoretically
\cite{hal.aps.100.31,lig.fra.3.1607,aub.mac.29.7261,aha.pha.204.1}.

(iii) A monolayer of a diblock copolymer is spread at the air-liquid
interface provided that one block is soluble in the aqueous subphase
while the other is hydrophobic
and acts as the 
grafting end of the chain
\cite{ken.jcp.103.2320,cao.far.98.1,kim.eur.24.229,gis.csa.86.171,ric.far,bij.lan.11.4467}. 
Similarly to the case of Langmuir monolayers
of short chains,
the grafting density of the diblock can be changed
continuously by moving a barrier on the air-liquid interface, hence changing
the surface pressure. Since the grafting density
is a controlled parameter,  different monolayer regimes can be tested
in one experiment offering a big advantage over 
polymer systems grafted or adsorbed on solid surfaces.

In the present work we address only the latter 
situation of a diblock copolymer
system at the air-liquid interface. Even here one can consider
two different cases as was realized experimentally.
In the first,
the block soluble in the liquid subphase has no surface activity.
Such a  case 
has been studied, e.g.,  by Kent et al.
\cite{ken.jcp.103.2320}.
In the second case, the block soluble in the liquid
subphase is attracted to
the air-liquid interface. This was 
realized experimentally by
a diblock copolymer composed of a
polyethylene oxide (PEO) and a polystyrene (PS) blocks. 
The PEO
homopolymer is soluble in water 
and is known  to adsorb spontaneously\cite{cao.far.98.1,kim.eur.24.229}
on the air-water
interface as can be
easily inferred from the  surface pressure isotherm showing
a 
``pseudo-plateau'' at medium surface pressures. 
Upon further increase in
the surface pressure,
the adsorbed PEO layer becomes
unstable. 
It is not possible to obtain a
very condensed surface layer since the monomers
can detach from the surface and dissolve in the bulk aqueous solution. 
However, if the PEO chain is linked chemically 
to another, water insoluble, chain (like a PS block),
further compression of the diblock copolymer into its
brush-like regime can be achieved\cite{gis.csa.86.171,ric.far}.

In a previous study of the PS-PEO copolymer\cite{bij.lan.11.4467},
different PEO chain lengths and a PS
block of 40 monomers have been used. 
A model based on a numerical version of
the self-consistent field (SCF) theory 
was proposed\cite{bij.lan.11.4467}
and took into account the adsorption
of the PEO monomers on the water surface. 
The model predicts in a 
qualitative way the experimental isotherms
for the PS-PEO monolayer and,  in particular, the existence of
a pseudo-plateau on the isotherm. From neutron
reflectivity data,  the progressive stretching of the chains in the
high surface density region is confirmed by fitting
of the results to the expected parabolic-shape concentration
profile \cite{mil.sci.251.905}. 

Although our study is mainly motivated by fundamental
issues related to pseudo two-dimensional polymer layer,
it is of relevance 
to important biomedical applications
where derivatives of PEO are used to modify 
adhesion properties of surfaces and interfaces.
We mention two examples.
In the first, grafting 
of polymers (in particular PEO with a hydrophobic group or block) 
to hydrophobic 
surfaces is a very promising way to
avoid non-desired protein adsorption onto specific surfaces
\cite{harris,szl.solid}. 
This type of surface modification
has great potential in increasing the biocompatibility of various 
materials. The second application relates to
grafting of PEO onto bilayers of liposomes. 
It has 
been shown that such a grafting increases the 
longevity of these liposomes in the blood stream,  making them 
practical vehicles for
enhanced drug delivery\cite{las.book}.
In the above two examples, the main function of the 
polymer layer is to form a steric barrier in
order to prevent the adhesion of proteins and other cells
in the blood stream. 

In the present work we  
further investigate experimentally pressure isotherms
of several PS-PEO systems
in the 
condensed as well as in the dilute and intermediate surface
density regimes. 
We use the
single chain mean-field (SCMF) theory to
calculate  pressure isotherms and  
concentration profiles of the experimental PS-PEO
copolymer system. 
The SCMF approach has been used successfully to
predict  the structural and thermodynamical properties of polymer
systems\cite{szl.review,car.mac.28.3197,szl.cocis.1.416}.
The predicted and measured
pressure isotherms are in very good agreement in the entire surface density
range. 
The only adjustable parameter, the adsorption energy of the PEO
monomers, is found to be roughly one 
$k_B T$. 
The SCMF calculation is complemented by a
mean-field approach in the low surface density regime, assuming
a two-dimensional layer of the PS-PEO copolymer.
Up to the plateau, there
is a very good agreement between the two theoretical approaches and also
with the experimental data.

Our paper is organized as follow:
in the Sec. II we will present our experimental technique. Section
III describes the SCMF theory. The experimental 
results on surface pressure isotherms for PS-PEO copolymers with various PS
and PEO block sizes and the predictions of the SCMF theory are 
presented in Sec. IV. The section ends with the mean-field calculation and
predictions. Finally, Sec. V 
is devoted to a discussion of the different results.

 \section{Materials and experimental technique}

 \subsection{Materials}

 We have used asymmetric diblock copolymers synthesized by sequential
anionic polymerization. The hydrophobic block consists of a polystyrene
(PS) chain having between 13 and 43 monomers. Each monomer
has a molecular weight
of 104 g/mole. The hydrophilic block is a polyethylene oxide (PEO) chain
that is OH terminated. The number of PEO monomers varies between 64 and
700, each having a molecular weight of 44 g/mole.  In the following, the
PS-PEO copolymer chains will be denoted as $\nps$--$\np$ where
$\nps$ is the number of PS monomers and $\np$ the number of PEO
monomers. The copolymers used in this work have a small polydispersity and
their properties are listed in Table 1. We note that 
the 31--$\np$ copolymers have
a partially deuterated PS block. However, this has no particular
importance for the present study. 

\subsection{Surface Isotherms}

The surface pressure isotherms have been obtained with a teflon Langmuir
film balance (Lauda -- FW 2) at $T=18^\circ$~C in a clean room.  The
maximum available surface  is 927 cm$^2$ and can be varied continuously
by moving a Teflon barrier. The surface pressure $\Pi$ is measured with a
Langmuir balance. The Langmuir film is obtained 
by depositing 
a small drop (with a micro syringe)
of about 100 $\mu$l of the copolymer in a
chloroform solution on the air-water interface.
The copolymer concentration is a few g/l.
 
Before compression, the film is allowed to equilibrate for 
about 15 minutes to
ensure full evaporation of the solvent and also 
to allow a re-adjustment of the
molecules. This re-adjustment is clearly more important for these long
chains than for short-chain surfactant molecules.  The compression
rate is kept constant at 55 \AA$^2$/molecule/min. We have checked that the
isotherms are unchanged when lower rates down to 2\AA$^2$/molecule/min are
used. As the copolymer chain length used here varies by a factor ten, we
adjusted accordingly
the speed of the barrier displacement 
in order to ensure the same compression rate (per monomer) in all the samples.

\section{Molecular Theory}
 
In this section we briefly describe the molecular approach that we use in
order to explain the experimental observations. The theoretical approach
is the single-chain mean-field (SCMF) theory which was originally
developed to treat packing of chains in surfactant aggregates
\cite{ben.jcp.83.3597} and later
was generalized for tethered polymer layers\cite{szl.review}.  
This theory has been widely
applied in a variety of polymer systems.  The predicted structural and
thermodynamic properties are in good agreement with experiments
\cite{ken.jcp.103.2320,szl.cocis.1.416}
and with
full scale Monte Carlo and molecular dynamics computer simulations
\cite{gre.binder}
Here we present a short derivation of
the theory concentrating on the points which are most relevant to the
PS-PEO monolayer. For an extensive discussion of the derivation and
application of the SCMF theory to tethered polymer layers, the reader is
referred to a recent review article\cite{szl.review}.

\subsection{Formalism}
 
The basic idea of the SCMF theory is to consider a central chain while
taking into account exactly all its {\it intramolecular} and surface
interactions, and treating the {\it intermolecular} interactions within a
mean-field approximation. The configurations of the central chain are
treated explicitly and the probability of the different conformations
changes as a function of the intermolecular (mean-field) 
interactions. Those latter
interactions depend upon the thermodynamic variables of the system, e.g.,
surface coverage and temperature. The probability distribution function
(pdf) of chain conformations is determined by minimizing the system
free-energy subject to the local incompressibility assumption. 
We note that the SCMF theory is different from the self-consistent
field (SCF) which was used in a previous study\cite{bij.lan.11.4467} 
of the PS-PEO system.
These differences are further discussed in Ref. \cite{szl.review}.
The SCMF theory can be applied in the entire range of surface coverage,
from the very dilute to concentrated coverages, and for different types of
solvents: good, bad and in {$ \Theta$} conditions. 

We model the PS-PEO layer at the air-water interface as two monolayers.
One is composed of PS segments and resides on the air side of the
interface. The other is the PEO layer residing on the water side. We
assume that the interface is sharp on  molecular length scales.  
While this is
not a necessary condition it simplifies the calculations without
affecting  the results in any
substantial way. 
It is, therefore, assumed that there is no coupling
between the PS and the PEO monolayers other than the fact that they both
have the same surface coverage. The theory presented below is 
applied for treating
a tethered polymer layer, with specific parameters chosen 
in order to represent the 
experimental PS-PEO copolymer layers as shown in Sec. IV. 

The central quantity of the theory is the probability distribution
function (\pdf) of chain conformations. From the knowledge of the \pdf\  any
desired conformational and thermodynamic averaged quantity can be
determined. The total free energy (per unit area) of the polymer layer 
can be written in terms of different entropy and energetic
contributions:
\be
\label{free}
\frac{W}{A} =-  T ( S_{c,pol} + S_{t,pol} + S_{t,solv}) +
\sigma\langle  \epsilon_{intra} \rangle   + 
\sigma\langle   \epsilon_{inter}\rangle +    
\sigma\langle   \epsilon_{s-p}\rangle 
\ee
where $T$ is the temperature.
The six
contributions to the free energy are listed below
and all depend
on averages with respect to
the \pdf\  of chain conformations, $P(\alpha)$. 

\begin{description}
 
\item[{\it i.}] 
{The conformational entropy of the chain molecules $S_{c,pol} =
-k_{B} \sigma \sum_{\{\alpha\}} P(\alpha) \ln P(\alpha) $, where $k_B$ is
the Boltzmann constant; $\sigma = {\cal N} / A =1/\Sigma$ denotes the
polymer surface density with ${\cal N}$ and $A$ being, respectively, the
total number of molecules and the area of the interface,
and $\Sigma$ the area per chain.
The sum runs over
all possible chain conformations $\{\alpha\}$.}
 
\item[{\it ii.}] 
{The translational entropy of the polymer chains, $S_{t,pol} =
- k_B \sigma \ln l^2 \sigma$, where $l$ is the polymer
segment length and $l^2$ is approximately the 
compact area per segment at the interface.}

\item[{\it iii.}] 
{The internal energy of the polymer-solvent mixture. One can
show that this energy can be written in terms of polymer-polymer
attractive interactions including both intra- and intermolecular
contributions as well as the surface-polymer (interface) interactions. 

The intramolecular contribution is $\langle \epsilon_{intra}\rangle
=\sum _{\{\alpha\}} P(\alpha) \epsilon_{intra} (\alpha)$.  The
energy $\epsilon_{intra} (\alpha)$ includes the sum of all non-bonded pairs
of segments, which are taken to interact
attractively  as the tail of a Lennard-Jones
potential. 
 
The surface-polymer interaction is $\langle \epsilon_{s-p}\rangle =
 \int_{0}^{\delta} \chi_{s-p} \langle n (z)\rangle \,\dz$, 
where $\chi_{s-p}$ is the strength of the surface-polymer
interaction. The average number of segments
of the central chain at distance $z$ from the interface
is given by
$\langle n (z)\rangle \dz=
\sum_{\{\alpha\}}P(\alpha)n(\alpha,z) \dz$, where $n (\alpha,z)\dz$
is the number of segments that a chain
in conformation $\alpha$ has at distance $z$ from the surface.
It is
assumed that all monomers within a distance $0 \le z \le \delta$ 
from the surface interact with the surface. This contribution
will turn out to be very important in understanding the experimental
PS-PEO pressure-area isotherms.

The intermolecular contribution has the form 
\be
\langle \epsilon_{inter}\rangle = \frac{1}{2\sigma l^4} \int \int \chi(|z-z'|)
\langle \phi_{p}(z)\rangle \langle \phi_{p}(z')\rangle \dz\dz'=
\frac{\sigma l^2}{2} \int \int \chi(|z-z'|)
\langle n(z)\rangle \langle n(z')\rangle \dz\dz'
\ee
where the
factor  $\frac{1}{2}$ corrects the double counting of the integral,
$\chi(|z-z'|)$ is the distance dependent van der Waals
interaction parameter (in units of $k_B T$),  and 
$\langle \phi_{p}(z)\rangle =\sigma l^3  \langle n(z)\rangle$ 
is the average
volume fraction of polymer in the layer defined between $z$ and $z+\dz$.}

\item[{\it iv.}] 
 {The translational entropy of the solvent molecules,
\par
$S_{t,solv}= -k_B l^{-3} \int \langle \phi_s(z)\rangle  
\ln \langle \phi_s (z)\rangle \dz$,
where $\langle \phi_s(z)\rangle$ is the average volume 
fraction of the solvent at distance $z$. 
The integrand is $z$ dependent due to the inhomogeneous
distribution of solvent close to the interface.}
 
\end{description}
 
A look at the different contributions to the free energy shows that the
intermolecular repulsive interactions 
have not been taken into account. To this end, we introduce packing
constraints that are $z$-dependent due to the inhomogeneous distribution
of solvent and polymer segments along the $z$ direction. The constraint
reads
\be
\label{constr}
\langle \phi_p (z) \rangle + \langle  \phi_s (z) \rangle=
\sigma l^3 \langle n(z) \rangle + \langle  \phi_s (z) \rangle   ~~ = ~~1
 ~~~~~ {\rm for ~ all ~ {\em z}}
\ee
where the first term is the volume fraction of polymer at distance $z$,
and the second is the volume fraction of solvent at $z$. 

In order to minimize the free energy density (\ref{free})
 with respect to the \pdf\  of chain conformations
and solvent density profile,
 subject to the packing constraint (\ref{constr}), we introduce a set
of Lagrange multipliers $\pi(z)$. This yields for the {\it pdf}
\begin{eqnarray}
P(\alpha) &=& \frac{1}{q} \exp \Bigg [ -\beta l^3 \int \pi (z)  n (\alpha,z) \dz
- \beta \epsilon_{intra}(\alpha) - 
\beta \int_0^{\delta} n(\alpha,z) \chi_{s-p} \dz
\nonumber \\
&& ~~~~~~~~~~~~~~~~ - \beta \sigma \l^2 \int \int 
\chi(|z-z'|) n(\alpha,z)
\langle  n (z') \rangle \dz \dz'  \Bigg ]
\label{pdf}
\end{eqnarray}
$q$ is a normalization constant 
ensuring $\sum_{\alpha} P(\alpha)=1$ and $\beta=1/k_B T$. Similarly, the
solvent density profile is given by
\be
\label{densprof}
\langle  \phi_{s} (z) \rangle = {\rm e}^{-\beta l^3\pi(z) - \beta\mu_{s}}
\ee
where $\mu_s$ is the chemical potential of the solvent molecules and it 
is constant at all $z$.

The only unknowns needed to determine the \pdf\  and the solvent density
profile are the Lagrange multipliers $\pi(z)$. 
They are obtained by substituting the
functional forms, Eqs. (\ref{pdf},\ref{densprof}), into the constraint
equation (\ref{constr}). In order to solve these equations, it is
convenient to discretize space into 
a set of parallel layers converting the integral
equations into a set of non-linear coupled equations. The details of the
calculation can be found in Ref. \cite{szl.review}. 
 
The physical meaning of the Lagrange multipliers can be seen from Eq.
(\ref{densprof}). These quantities are related to the osmotic pressure
necessary to keep the chemical potential of the solvent constant at all
$z$.  Furthermore, they represent the effective repulsive interactions
(lateral pressure) between the polymer that force the chains to stretch
out from the surface into the bulk solution. 

In order to compare the predictions of the theory with the experimental
pressure-area isotherms we need to derive theoretically
the lateral pressure. 
This is obtained by a differentiation of the total free energy $W$ with
respect to the total area $A$, i.e. 
 \be
\label{formalpre}
\Pi =-\left (  \frac{ \partial W }{\partial A} \right )_{{\cal N},T}
\ee
 Taking the derivative of the free energy (\ref{free}), after substituting
the \pdf\  and the solvent density profiles, Eqs. (\ref{pdf},\ref{densprof}),
yields the following surface pressure
 \be 
\label{press} 
\beta \Pi = \sigma  + \int \beta \pi(z) \dz - \sigma  N +
\frac{1}{2 l^{4} } \int \int \beta \chi(|z-z'|)
 \langle \phi_p (z) \rangle \langle
\phi_p (z')\rangle \dz \dz' 
\ee
where $N$ is the tethered polymer chain length (either the PS or PEO blocks). 
The lateral pressure is 
readily calculated once the Lagrange multipliers are determined. 

\subsection{SCMF Parameters}

The determination of the osmotic pressure profile $\{\pi(z)\}$ requires
the solution of the constraint equations. The needed parameters are:
(i) the
surface density of polymer $\sigma$ (or equivalently, the area per chain,
$\Sigma=1/\sigma$),
(ii) the strength of the polymer-polymer
attraction, $\chi(|z-z'|)$ 
(or, alternatively, the Lennard-Jones potential
strength), (iii) the strength $\chi_{s-p}$ and the width $\delta$ of the
polymer-interface interaction range, 
and (iv) the set of single chain conformations of
the polymer under study. 

We have chosen to use the rotational isomeric state model for the polymer
chains. For both types of chains, PS and PEO, each segment is modeled as
having three possible states. The three states are taken to be
iso-energetic and each segment represents a styrene or an ethylene oxide
(EO) monomer. The internal structure of the monomers, i.e., detail
chemical structure and internal energy of the different states, is
replaced by an effective monomer length representing the bond between the
neighboring monomers. The length of the segments is $\lps = 4$\AA\ and
$\lpeo=3$\AA\ for the styrene and EO units, respectively. Note that these
values compare reasonably well 
with the proper length of the chemical segments
in each type of polymer. 

The constraint equations are solved by discretizing space into parallel
layers of thickness $\delta=1.84~ \lpeo$. This value of $\delta$ has been 
found to be the more convenient in the calculation, however the predictions
of the theory are independent of the choice of $\delta$\cite{szl.review}. 
This transforms the integral
equations, Eq. \ref{pdf}, into a set of non-linear equations which are
solved by standard numerical methods. We remark that once the set of single
chain conformations is generated, it is used for all the calculations for
different surface coverages and conditions. Further technical details of the
calculations can be found in Ref. \cite{szl.review}.

The values of the parameters defined above have 
not been optimized in order to
obtain the best possible agreement with experiment.  They were simply
chosen on the basis of reasonable physical values. 
The value of the Lennard-Jones strength is $0.33k_B T$
for the styrene-styrene and
$0.9k_B T$ for the EO-EO attractions, respectively.
 The only parameter whose value is not known experimentally 
is the EO-interface attraction parameter, $\chi_{s-p}$.
A few values of this parameter
have been checked (in steps of $0.5~k_{B}T$). 
The one we finally used, $\chi_{s-p}=-1.0k_B T$, fits best the
experimental data. We remark that
even 
better agreement with the experimental observations
can be achieved by further optimization of the parameters,
 but this is left to
future studies.

\section{Experimental Observations and Theoretical Results}

\subsection{Isotherms and the PEO Layer Structure}

On Fig. 1 we plot the experimental isotherms corresponding 
to different PEO chain
lengths while keeping the length of the PS block fixed at $\nps=31$ (in one case
$\nps=30$).
A pseudo-plateau in the pressure-area
isotherms can be seen in Fig. 1 signaling a small
pressure variation over a large change in the area per molecule
($\Sigma$), especially for the longer chain lengths. This is in accord
with previous experiments of Ref. \cite{bij.lan.11.4467}.
The pressure-area isotherms in the plateau and the highly compressed regions
are completely reproducible, for both compression and decompression cycles
as has been already discussed in Ref. \cite{bij.lan.11.4467}.

Hysteresis effects are observed
only for the diluted region. Furthermore, for monolayers spread out at
molecular areas a little bigger than the plateau values, the first
compression is different from the following ones. For such a deposition we
suspect that the chains are entangled at the surface without being able to
disentangled during experimental times. Another problem with deposition
done for $\Sigma$ close to the plateau value is the appearance of two
plateaus instead of one. For this reason, all experiments have been
performed after spreading out the monolayer at very low pressure (and high
$\Sigma$).  The pressure isotherms have been always measured while
compressing the monolayer. A hysteresis is still present under these
conditions and leads to smaller pressures 
at large
molecular areas during decompression . This could be explained by 
pre-orientation during the
decompression: the entanglement of the chains at the interface is
less pronounced when a polymer ``brush" is decompressed than when it is
compressed from a two-dimensional monolayer. Hence, it leads to smaller
surface pressures in the former case. 

Figure 2 shows the prediction of the SCMF theory 
together with the experimental
pressure-area isotherms.  As the SCMF method is
limited by chain length of about 200 monomers, we show isotherms
for three different chain lengths up to this value. The 
theory is in very good agreement with the experimental observations. The
presence of the plateau is a manifestation of the preferential adsorption
of the EO monomers at the air-water interface, as earlier suggested by
Bijsterbosch et al. \cite{bij.lan.11.4467}.

In order to obtain quantitative agreement
between the measured isotherms and the theoretical 
findings it is necessary to
take into account the role of the PS block
(see Sec. IV.B below). The short PS block resides on
the air side of the interface, is not attracted to the air-water interface
and has strong attractive interactions with monomers of its own type.
Namely, the PS is in a poor solvent conditions (the poor ``solvent" being
the air). In addition,
the strength of the attraction between the EO monomers and the interface
is found to be 
considerably larger than the estimates given by Bijsterbosch et al. 
\cite{bij.lan.11.4467}. The
value for $\chi_{s-p}$ 
we used and which fits the experiments is about $-k_{B}T$ per
monomer. We have also performed calculations for bigger values of the
EO-interface attraction in order to check if there is a first-order phase
transition as predicted by Ligoure
for long polymer chains\cite{lig.fra.3.1607} 
However, our calculations 
as applied to intermediate chain lengths, do not show
any indication of such a first-order transition even for attractions per
monomer as strong as $\chi_{s-p}=-4k_{B}T$. 

As the monolayer is compressed,
starting from very large $\Sigma$, the  isotherm shows a 
relatively sharp increase of
pressure as $\Sigma$ is decreased, 
followed by a plateau. 
The term ``plateau" is loosely used hereafter (and by others)
to denote  a region for relatively small $\Sigma$
where the pressure reaches a pseudo--plateau and
increases only gradually as the layer
is compressed. Upon even further compression, 
there is a very large increase of the
pressure. This is due to the very large densities of the chains at small
$\Sigma$ resulting in very high pressures for all types of
solvent qualities. 

The understanding of the origin of the plateau is intimately related to
the structural behavior of the polymer layer as a function of the area per
molecule. We consider only the PEO part of the layer which is responsible
for the presence of the plateau and for the overall shape of the isotherm.
The PS block ``tunes" the pressure in the different regimes but does not
change the qualitative shape of the isotherms (see Sec. IV.B below).

Figure 3 shows the pressure as a function of $\Sigma$ (the area per
molecule) and $\langle \phi_p(z)\rangle$, the
density profiles of the PEO chains, for four
different  surface densities, as calculated from the SCMF theory. 
 The density profile at large $\Sigma$ corresponds to distances
between tethering points larger than the radius of gyration of the chain.
The density profile reveals that at these large $\Sigma$ values, the PEO
chains form a quasi two-dimensional layer at the air-water interface due
to the strong monomer-interface attraction. 

Up to the onset of the plateau in the pressure-area isotherm, as $\Sigma$ is 
decreased, the density profile looks very similar to the very
dilute case. Namely, down to the plateau 
region of the isotherm, the PEO forms a quasi
two-dimensional
layer. Comparison of the increase of the pressure with surface density in
this low density region (high $\Sigma$)
with that of non attracting surfaces 
reveals that
the increase in the former case is
much sharper than in the non-attracting case. The reason is that when
the monomers are attracted by the interface, 
their  local density becomes higher
resulting in larger lateral monomer repulsion and hence
larger pressure. 
This is to be contrasted with non-attractive interfaces where the
chains stretch quite substantially into the bulk solution (the
``mushroom" regime). 

The appearance of the plateau in the pressure-area isotherm corresponds to the
point where the density profile shows that a high fraction 
of the PEO monomers is adsorbed on
the surface. Another feature of this plateau,
as seen from the density profiles, is the appearance of a second
structure reminiscent to a brush 
composed of chains with the same number
of segments as the PEO chains but without the ones attracted to the surface.
This shape of the profile remains basically the same as $\Sigma$
decreases.  Upon further compression, the brush-like part of the
profile corresponds to effective longer chain lengths because the polymer
density at the interface does not change.

The formation of the adsorbed layer at low surface coverages implies that
for this region the pressure should exclusively be determined by the number of
segments in the layer. Therefore, the pressure should scale with $\Sigma/\np$. 
Figure 4 shows that for all $\Sigma$ values
higher then the onset of the plateau, the
pressure scale reasonable well with $\Sigma/ \np$
both as measured experimentally for $\nps=30$ (4a) and  
from the SCMF calculation (4b). 
In addition, it was checked experimentally for 
isotherms with $\nps=13$ and 43.

Clearly, once the brush appears the pressure
shows a behavior  combining the highly dense adsorbed layer one with
that of the short chain length brush. Empirically, we find that in the
high surface coverage regime the calculated pressure scales with 
$\sigma (\np)^{0.6}$ as
shown in Fig. 5. However, as discussed in detail in 
Refs. \cite{car.mac.28.3197,szl.review},
special care
is needed in trying to describe these polymer layers with scaling concepts
since the chain length are too short and there is no easy definition of
the scaling regimes such as the ``mushroom" and ``brush" regimes. 

From the experimental observations and the SCMF predictions we find 
that the width of the plateau region increases with the
size of the PEO chain, while the height of the plateau is independent of
$\np$ and is equal to about 10 mN/m. The 
height of the plateau is a function of the  PS chain length (see below)
and from the SCMF calculations (not shown) it also depends on the affinity 
of the interface to the EO monomers (our $\chi_{s-p}$ parameter).

Figure 6 shows the PEO contribution to the pressure 
$\Pi$, and the square of the volume
fraction of the adsorbed layer $\langle \phi_p(1)\rangle^2$,
including segments at 
distances $0\le z \le \delta$ from the surface,
as a function of $\Sigma/\np$, as calculated from the SCMF theory.  
For $\Sigma$ higher than the onset of the plateau,
the square of the volume fraction of the adsorbed
layer is identical in shape, and, therefore, fully determines, the
pressure of the PEO layer. This is again a manifestation of the 
quasi two-dimensional behavior of the PEO layer up to the onset of 
the plateau.
For higher surface densities (lower
$\Sigma$), the polymer layer starts to behave like a combination of
a 2d layer and a 
brush.  Hence, there
is no reason there why $\Pi$ and $\langle\phi_p(1)\rangle^2$ 
will be proportional.

\subsection{The Effect of PS Block}

The effect of the PS chain length, $\nps$,
on the pressure-area isotherms are
presented in Fig. 7a for the experimental isotherms and in Fig. 7b for the 
SCMF predictions. 
As explained above, the pressure isotherms are more
conveniently plotted 
as a function of $\Sigma/\np$.  
We first note that the dependence of 
the isotherms on $\nps$ is rather small for 
$\Sigma$ values higher than the plateau region. 
However, When $\nps$
increases, the height and the slope of the plateau decreases. The experimental
isotherm of a pure PEO homopolymer is plotted (triangles)
on the same figure for comparison. 
For the latter, a
plateau is observed as well
but at a lower surface pressure. 
There is no steep rise in the pressure at small molecular area because
the absence of an anchoring hydrophobic block.
Instead the pure PEO layer loses its stability. 

For very short PS chains, the PS block may be thought of as being in a
liquid ``collapse" state (bad solvent conditions).  In fact, to the best
of our knowledge, no previous measurements exist for the glass transition
temperature, $T_g$, of PS in a diblock copolymer confined to a
two-dimensional surface. In bulk (three-dimensional) systems, $T_g$ is
equal to 23$^\circ$, 67$^\circ$ and 
76$^\circ$ C, respectively for $\nps=$ 13, 30 and
43.
Careful surface viscosity experiments are needed in order to further
clarify this point. 

The SCMF 
results shown in Fig. 7b are in good agreement with 
the experimental observations on the effect of the PS block. 
The PS block is modeled in the theory as being in poor 
solvent conditions so that without being linked to the PEO block,
the PS layer
would undergo a two dimensional``gas-liquid" (collapse) phase transition. 
However,
the presence of the attached PEO block (for which the water is a good
solvent), ensures that the PS layer is stable for all surface
coverages down to very compressed states.

\subsection{Mean-Field Theory: Dilute Region}

In order to complement the numerical and more quantitative results
of the SCMF theory, a simple mean-field approach
is proposed. Although it is more qualitative, it offers a simple
explanation of the PS-PEO isotherms for the dilute regime 
and is in semi-quantitative accord both with the experiments 
and the SCMF approach.

Based on the experimental isotherms and SCMF polymer profiles, we
model the polymer layer in the 
dilute region of the isotherm 
($\Sigma$ larger than the plateau
value), as a polymer layer 
where $\npt$ monomers of each PEO chain
form an effective 2d layer whose thickness does not vary upon
compression.  Since we expect $\npt\approx\np$ 
for densities smaller
than the 2d overlap density, 
the effect of the other PEO monomers 
which do not participate in the 2d layer
($\np-\npt$ per chain) is neglected.
On the other hand, the 
attraction of the PEO monomers to the air-water interface plays
an important role in the 2d behavior of the PS-PEO copolymer layer.

Let us first  consider the thermodynamics of a 2d
PEO layer without the effect of the PS chains. They will be included
later. The free energy of mixing can be written as a 2d Flory-Huggins
free energy in terms of the area fraction occupied by the PEO
monomers on the surface: $\phi_2=l^2\npt\sigma=l^2\npt/\Sigma$. 
This area fraction
corresponds to the volume fraction of the polymer chains in the first layer,
$\langle\phi_p(1)\rangle$ as was defined earlier (Sec. III.A).

\be
\label{d1}
F_{\rm site}~~=
~~\frac{l^2}{A}{W}~~=
~~{k_B T}\left[ \frac{\phi_{2}}{\npt}\ln\phi_{2}~~+
~~(1-\phi_{2})\ln(1-\phi_{2})\right]
+\chi_0\phi_{2}(1-\phi_{2})+\chi_{s-p}\phi_{2}
\ee
where $l=\lpeo\simeq 3$\AA\  is
the EO monomer size. 
The first two terms are just the entropy of mixing of the PEO monomers
at the interface, the third term is the enthalpy of mixing, and the last  
term represents the attractive surface interaction of the PEO monomers with
an interaction parameter $\chi_{s-p}$ varying between
$-0.5k_B T$ and $-1.0 k_B T$.

The PS block is anchored at the surface on the air side. As air acts
as a bad solvent for the relatively short PS chains, the chains can be
considered to be in a collapse state, each occupying an area
$\Sps=(\nps)^{2/3}\lps^2$ which is estimated around $\Sps
\simeq 150 - 200$\AA$^2$ for $\nps\simeq 30-40$ and 
the PS monomer size is taken as $\lps=4$\AA.
We take into account two effects of the PS block on the PEO
chains. First, the area available for the PEO monomers is
reduced.
The available area per PEO chain is only
$\Sigma-\Sps$. Second, we take into account the attractive
PS-PS interactions.

Altogether the more general free energy (per site) is

\be
\label{d3}
F_{\rm site}~~=~~
{k_B T}\left[\frac{\phi_{2}}{\npt}\log\phi_{2}~~+~~(1-\phi_{2})\log(1-\phi_{2})
\right]
+\chi_0\phi_{2}(1-\phi_{2})+\chi_{s-p}\phi_{2}+\chi_{1}\phips(1-\phips)
\ee
where
\be
\label{d4}
\phips=\Sps/\Sigma  
~~~~~~~~~~~\phi_{2}=l^2\npt/(\Sigma-\Sps) 
\ee

The surface pressure can be calculated by differentiating the total free
energy $W=A F_{\rm site}/l^2$ with respect to the total surface area $A$.
Alternatively, the surface pressure $\Pi$ can be obtained directly
from the free energy density $F_{\rm site}$, yielding
\be
\label{d5}
\Pi=-\frac{1}{l^2}\frac{\partial(\Sigma F_{\rm site})}{\partial\Sigma}=
-\frac{1}{l^2}(F_{\rm site}+\Sigma \frac{\partial F_{\rm site}}{\partial\Sigma})
\ee

By estimating the values of the parameters: $\npt$, $\chi_0$, $\chi_{1}$,
$\Sps$ and $\chi_{s-p}$, we can obtain $\Pi(\Sigma)$ from Eqs.
(\ref{d3})-(\ref{d5}). However, we present here only the analytical
expansion of the pressure $\Pi$ in powers of $\Sigma^{-1}$ valid for the
dilute limit, $\Sigma \gg \Sps\simeq 200$\AA. 
Neglecting the entropy of the PEO
chains (since $1/\npt$ is rather small) and expanding Eq. (\ref{d3}) we get

\be 
\label{d6} 
\frac{F_{\rm site}}{k_B T} ~~\simeq~~ 
(-1+\beta\chi_{s-p}-\beta\chi_0)\phi_0 +
(-1+\beta\chi_{s-p}+\beta\chi_0)\phi_0\phips
+\frac{1}{2}(1-2\beta\chi_0)\phi_0^2+\beta\chi_{1}\phips-
\beta\chi_{1}\phips^2
\ee 
where $\phi_0=l^2\npt/\Sigma$ is the unperturbed PEO area fraction
without the effect of the PS block and $\beta=1/k_B T$. 
The pressure is readily
calculated from (\ref{d6}) to be

\be
\label{d7}
\frac{l^2}{k_B T} \Pi~~\simeq~~
\left[(-1+\beta\chi_{s-p}+\beta\chi_0)\frac{\Sps}{l^2\npt}
-\beta\chi_{1}\left( \frac{\Sps}{l^2\npt}\right)^2 
+ \frac{1}{2} - \beta\chi_0\right]\phi_0^2
\ee

Taking the following estimates for the parameters: $\chi_0\simeq 0$,
$\Sps/l^2\npt \simeq \lps^2\nps^{2/3}/l^2\npt \simeq 0.1$, we 
get the 
following estimate of the leading term in the pressure as function
of the surface coverage $\sigma=1/\Sigma$

\be
\label{d9}
\frac{\Pi}{k_B T} ~~\simeq~~ 
\left[\frac{1}{2}-0.1(1-\beta\chi_{s-p}) 
-0.01\beta\chi_{1}\right](\npt l\sigma)^2 
\ee

From (\ref{d9}) we see that the interaction effect of the PS monomers is
rather small (for the dilute regime). For reasonable estimates of the
PEO-surface interaction parameter $\beta \chi_{s-p}\simeq -1.0$ we get that the
pressure depends on $(\npt l\sigma)^2$ with a prefactor of 0.3. This
result is in accord with the SCMF calculation presented on Fig. 6
where it is seen that indeed $\Pi\sim\phi_2^2\sim(\sigma \npt)^2$
for all dilute surface
densities till the plateau. Note that
$\phi_2$ as was defined in this section corresponds to 
$\langle \phi_p(1)\rangle$
of the SCMF calculations. 
 Hence, the simple calculation presented in this
section gives a simpler description
(albeit more approximated) to
the thermodynamics of the PS-PEO layer for compression till the plateau
values. We note again that the PEO effective size is taken as $\npt$ and
includes only those PEO monomers that lie on the 2d surface. 

Equations (\ref{d7})-(\ref{d9}) can be compared with the experimental
isotherms. To do so we plot the experimental isotherms $\Pi/\np^2$ as a
function of $1/\Sigma^2=\sigma^2$ for the various PS-PEO copolymers. For
surface densities smaller than the plateau values, we observe the expected
linear dependence for all our copolymer systems. From the slope given by
(\ref{d9}) $0.3 l^2 \npt^2/ (k_B T\np^2)$, the fraction of the PEO
monomers adsorbed at the interface is deduced assuming that the size of a
PEO monomer is $l=3$\AA.

On Fig. 8, the variation of this fraction $\npt/\np$ is presented as a
function of the chain length $\np$. For short chains, the fraction of
monomers adsorbed at the interface is less than one and increases with
$\np$. It reaches an asymptotic value for the longer chains. Note that
this asymptotic value is larger than one, due probably to the
approximations used in determining the prefactor. We can consider that
part of the PEO monomers are constrained by their link to the PS block and
effectively are shielding this block from a unfavorable contact with the
water. The total number of monomers affected by this constraint should be
independent of the length of the PEO chain. Consequently, their ratio to
the total chain length should decrease with $\np$ and will be negligible
for the long chains.

%
\section{Summary}

The structure and thermodynamics of PS-PEO layers have been 
studied using a combination of experimental observations, molecular
theory (SCMF) and a simple mean-field approach. The picture that emerges  
is that at low surface coverages the PEO forms a two dimensional layer
due to the attractions of the EO segments to the air-water interface. This can
be understood since
the ethylene oxide monomer shows amphiphilic (surface activity)
behavior. Upon further
compression, the adsorbed polymers
start to overlap and 
there is a plateau in the pressure isotherms. The plateau 
corresponds to the region where there is a 2d layer of constant density of
ethylene-oxide and a water soluble brush-like layer. At very high surface
densities the pressure increases very sharply due to the strong repulsion
between the stretched and adsorbed parts of the layer. The PS block 
is in a poor solvent environment, hence being in a collapsed state 
at all surface 
coverages.

The most important difference between the PS-PEO 
and other block copolymer systems is the attraction of the EO monomers
to the interface. 
The ramifications of this EO attraction to hydrophobic
surfaces is very important in the understanding and design of
biocompatible materials and polymer decorated liposomes.

The structural changes of the PEO layer are the result of the competition
between the solubility of the EO monomers in water and their favorable 
energy of interaction with the interface. This energy is estimated from the 
SCMF to be or order of
one $k_{B}T$. This indicates that the PEO layer cannot be treated
within the general framework of ``mushrooms'' and ``brush'' regimes usually
considered for layers of tethered and long polymers.
A theoretical description of this
layers requires the explicit consideration of the 2d EO layer formed at
the interface for all polymer surface coverages.

The role of the PS block is to anchor the polymer to the interface and to 
tune the values of the pressures, i.e. the height of the plateau region. This 
is because the chain lengths of the PS blocks studied here are short enough,
as compared to the PEO block, such that the overall behavior of the layer is 
determined by the water soluble block (PEO). 

\bigskip\bigskip
\bigskip\bigskip

\noindent{\it Acknowledgments}

One of us (IS) wishes 
to acknowledge support from the  NSF under grant No. CTS-9624268, while DA
acknowledges support from US-Israel Binational Foundation (BSF)
under grant No. 94-00291. IS is a Camille Dreyfus teacher-scholar.



\pagebreak

\section*{Table Caption}
 
\begin{description}

 \item[Table 1]{The different copolymers used in the isothermal
measurements. $\nps$ is the
number of PS monomers and $\np$ 
the number of PEO monomers. The copolymers
labeled with D have a PS block which is partially deuterated.}
 
\end{description}

\bigskip\bigskip

\section*{Figure Captions}
 
\begin{description}

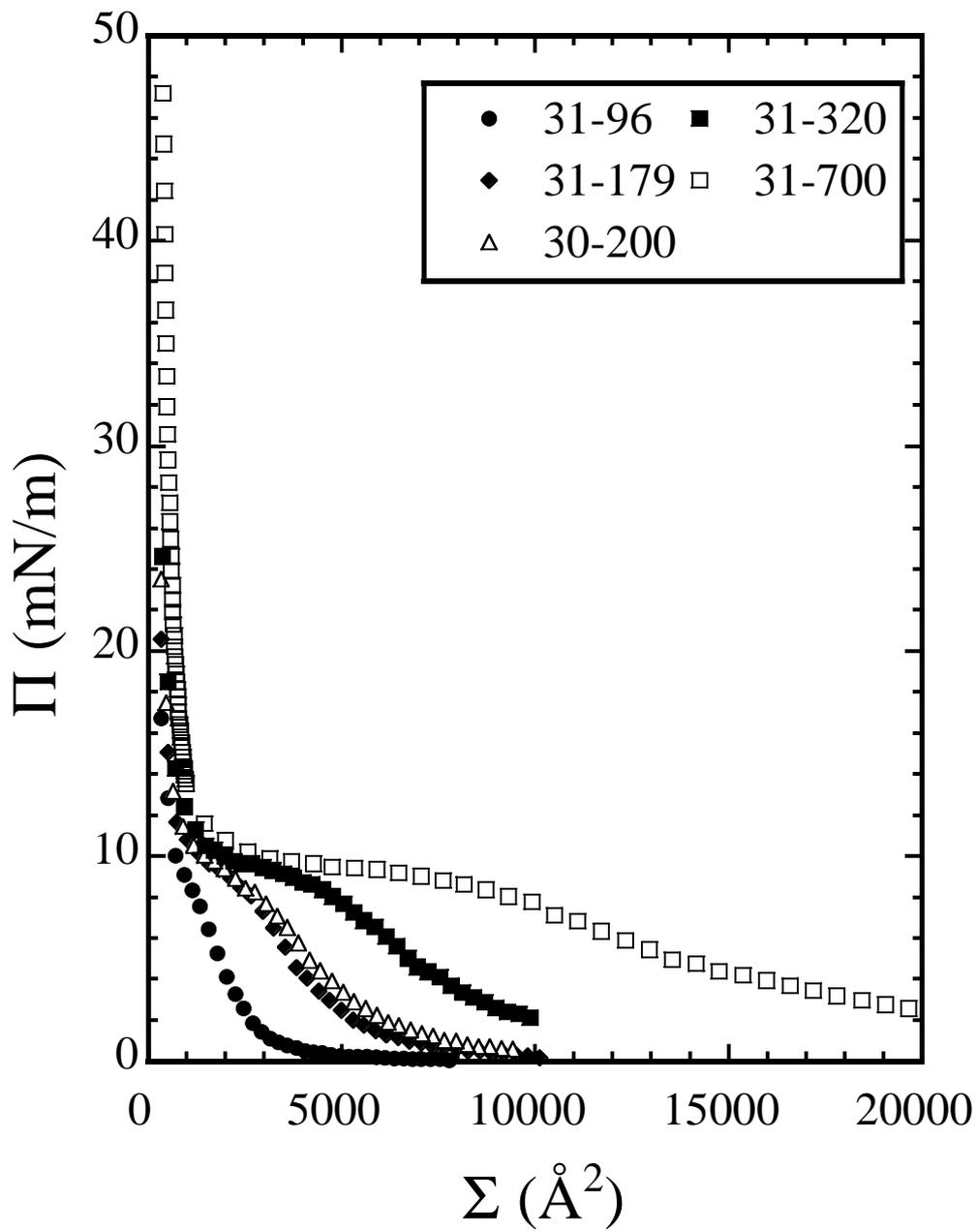
\item[Figure 1]{Experimental surface pressure -- area isotherms at 
$T=18^\circ$ C of PS-PEO copolymers with several 
sizes of the PEO block and with the same PS block, $\nps=31$, beside
one case with $\nps=30$. (We will ignore the difference between these
two similar PS sizes).}
 
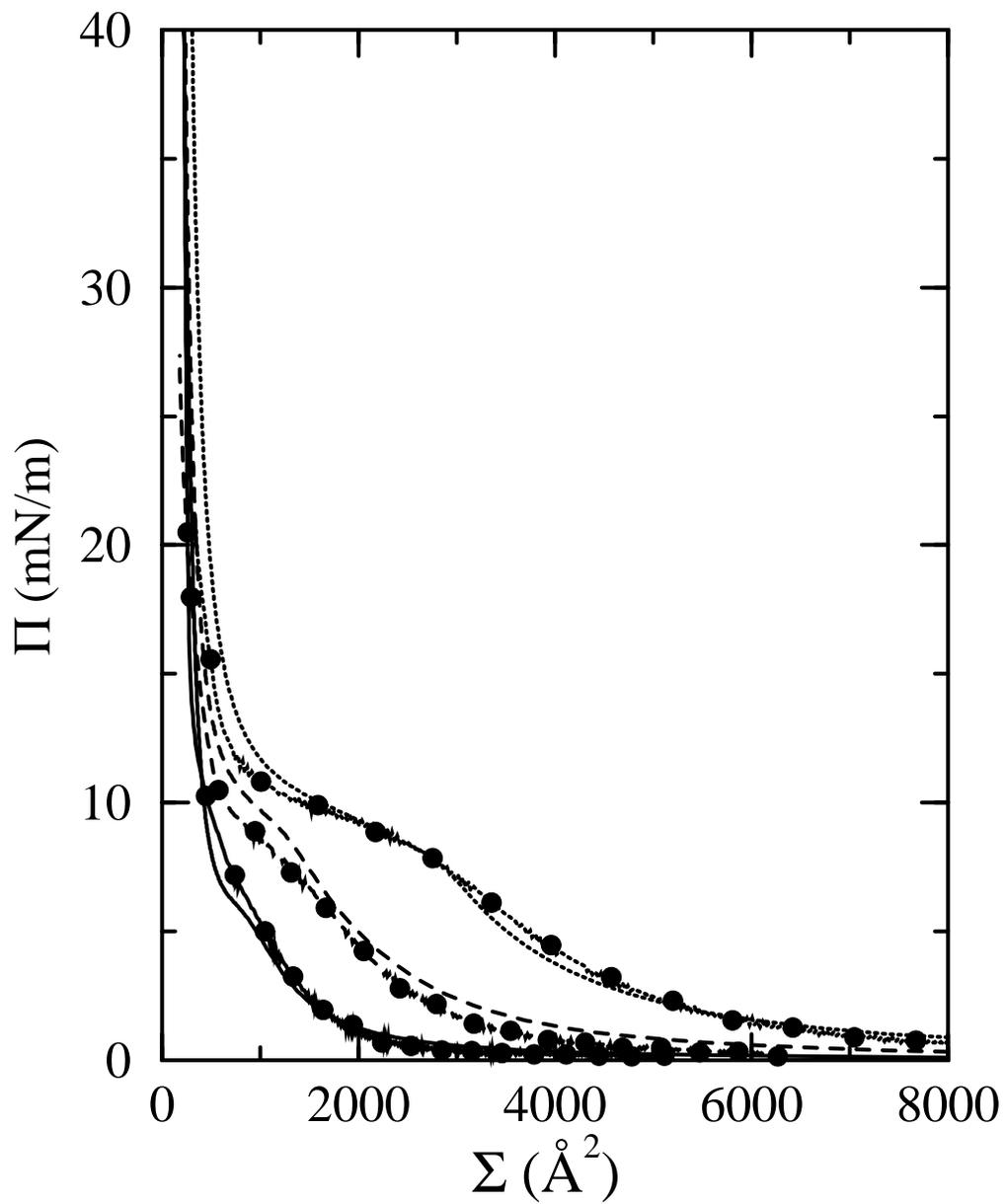
\item[Figure 2]{direct comparison of experimental pressure-area isotherms
with those obtained from the SCMF theory, for three different copolymers:
30-179 (dotted line); 30-100 (dashed line); 43-64 (smooth line).
The experimental curves are denoted by their data points (full circles) 
and are very close to the calculated ones.}

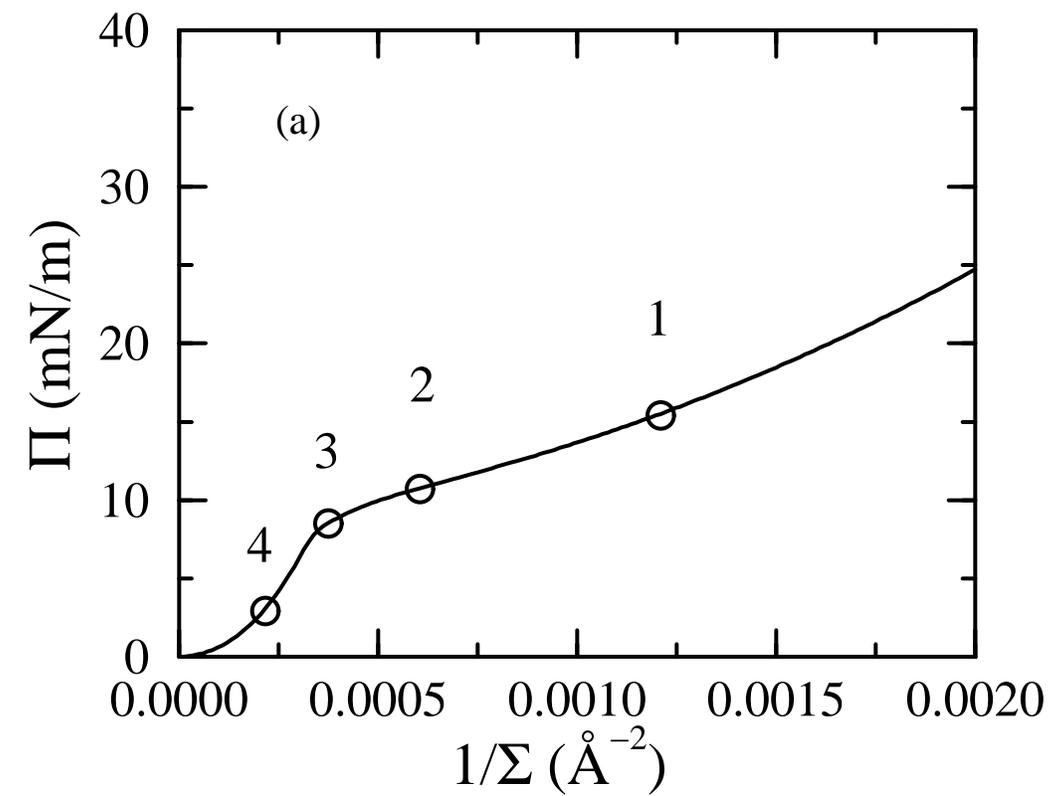
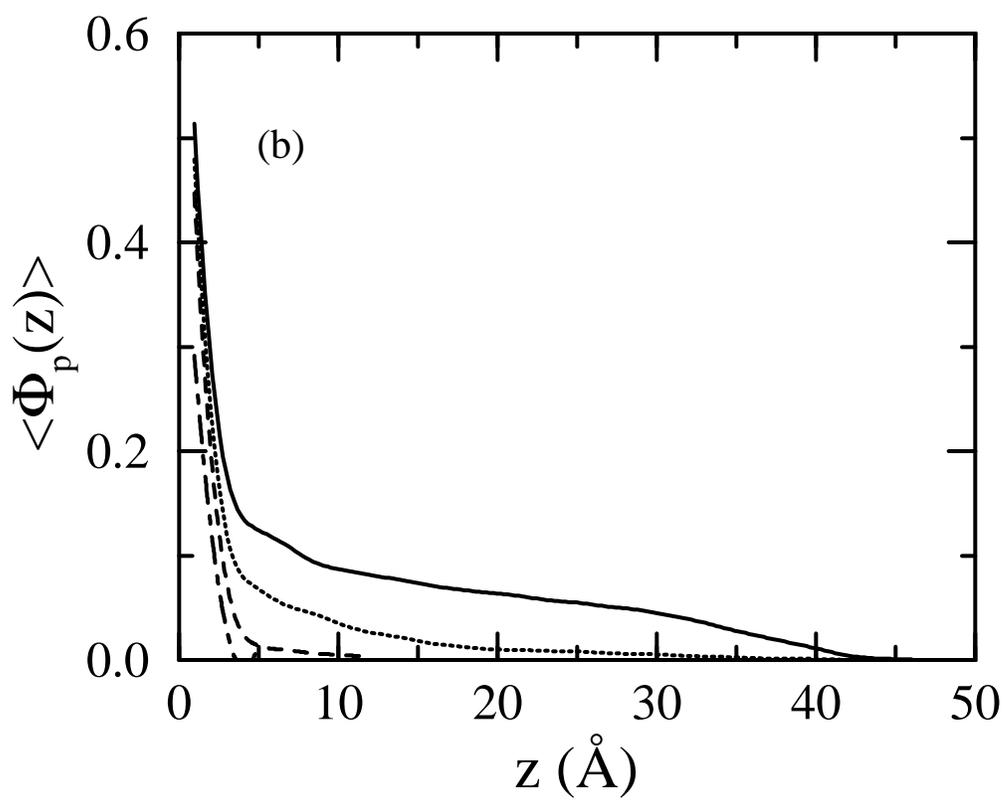
\item[Figure 3]{(a) Pressure isotherm as function of
the surface density $\sigma=1/\Sigma$; (b) density profiles 
$\langle \phi_p(z) \rangle$ 
as function of the distance from the air-water
interface, $z$; both  obtained from the SCMF
theory. Only the contribution of the PEO block is 
shown. The smooth, dotted, dashed and dashed-dotted
lines in (b) corresponds, respectively, to the points marked
as 1, 2, 3 and 4 on the 
isotherm in (a) having the surface density: 
0.0012, 0.0006, 0.00038 and 0.00022 \AA$^{-2}$,
respectively.}

\item[Figure 4]{(a) same as Fig. 1 but 
the data is plotted as a function of the 
area/molecule rescaled by the number of PEO monomers $\np$: $\Sigma/\np$.
(b) The contribution of the PEO block
to the pressure calculated from the SCMF theory 
as a function of the same rescaled area as in (a) for three PEO block 
sizes: $\np$=64 (smooth line); 118 (dotted line); 179 (dashed line).}

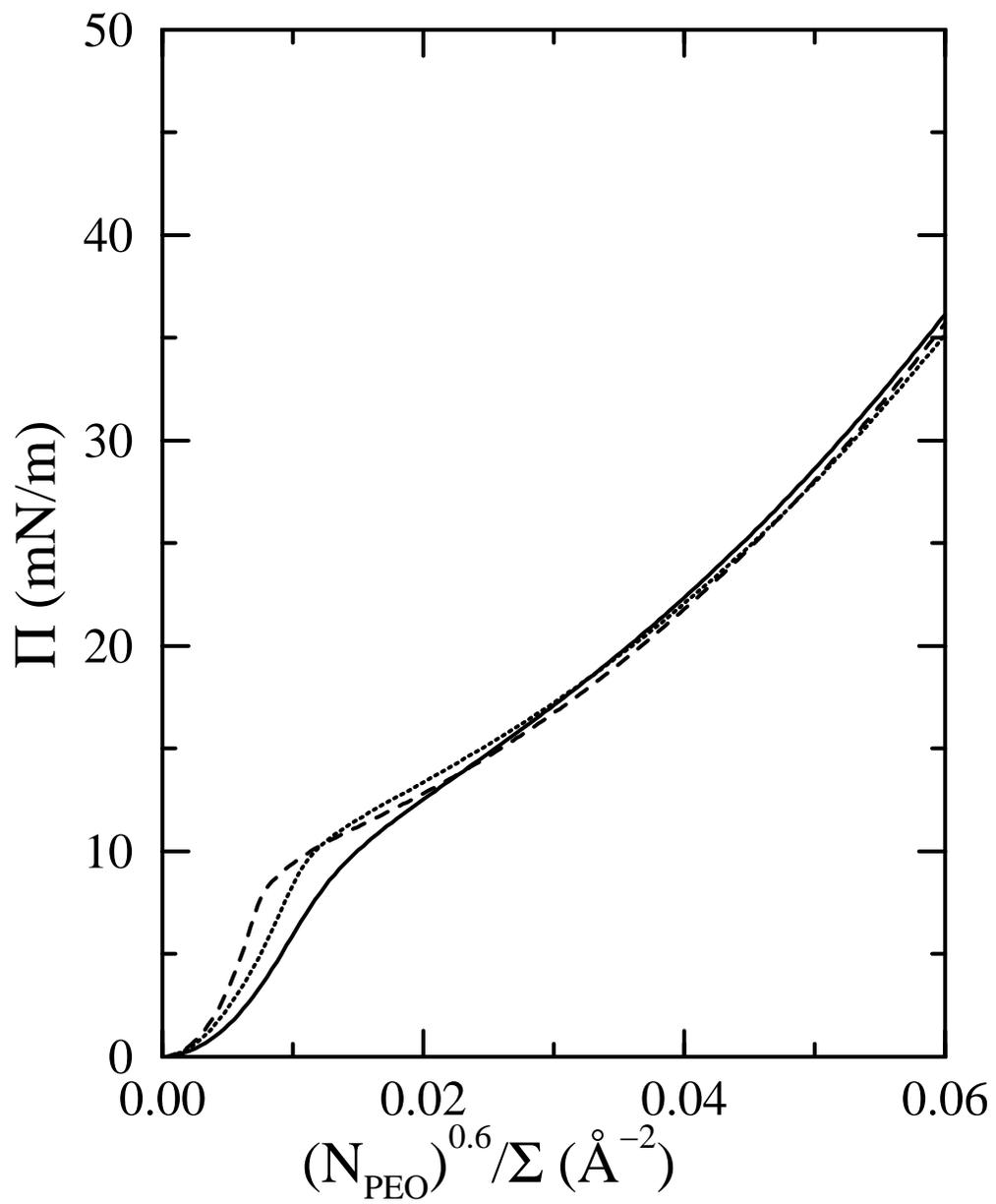
\item[Figure 5]{Same values and symbols as
for the isotherms in Fig.~4b but the isotherms here
are plotted with a rescaled surface density,
$(\np)^{0.6}/\Sigma$, in order to show the scaling behavior in the 
``brush" regime (high $\sigma$).}

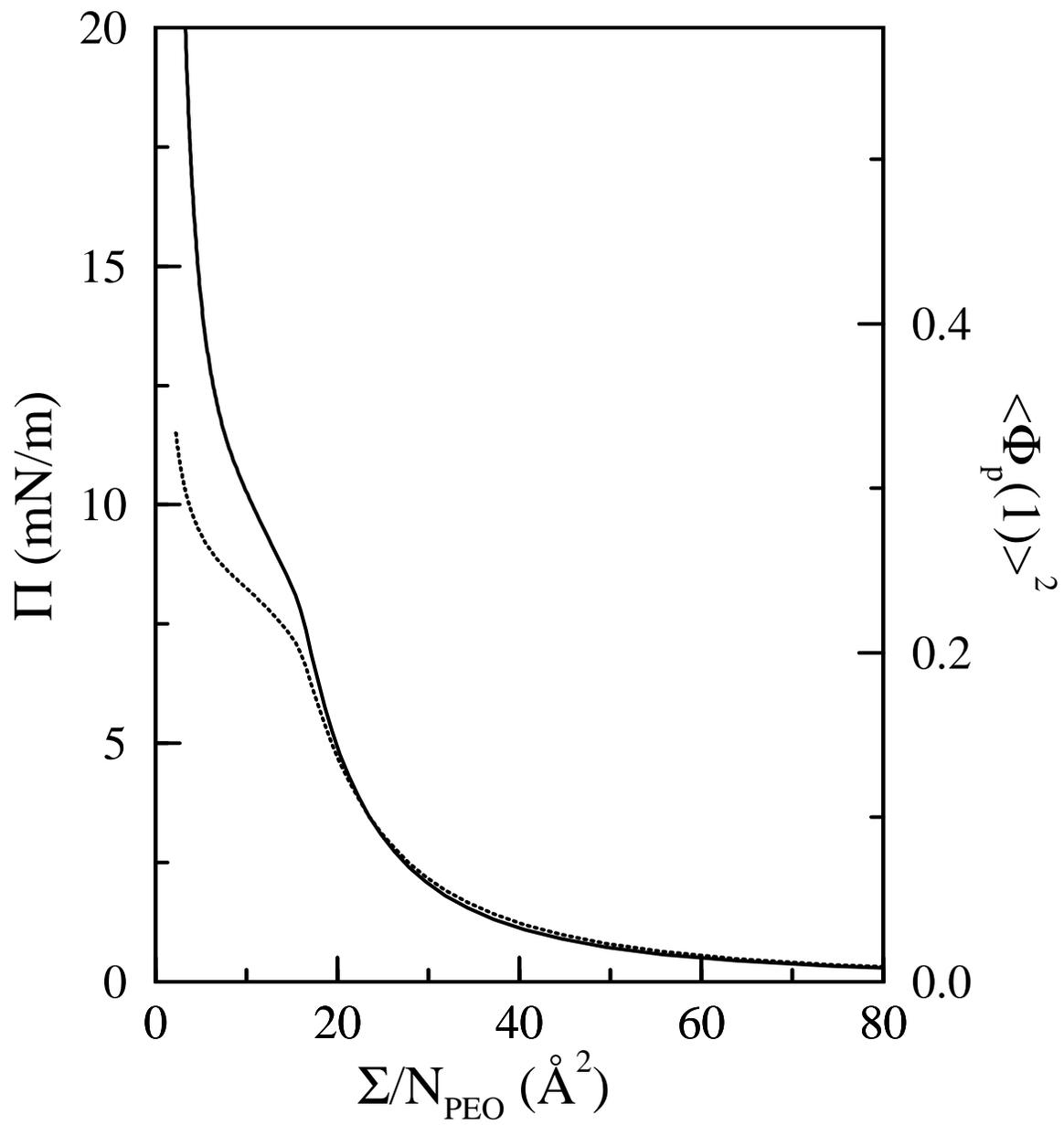
\item[Figure 6]{Calculated contribution of the
 PEO block to the surface pressure 
(smooth line) and $\ 25 \langle \phi_p(1)\rangle^2$ (dotted
line) as function of $\Sigma/\np$ for $\np=179$. The two
curves become identical in the large $\Sigma$ region (roughly for
$\Sigma/\np\ge 18{\rm \AA}^2$) demonstrating 
that $\Pi$ is proportional to
$\langle\phi_p(1)\rangle^2$ in this region. 
This essentially two-dimensional behavior
is expected for the dilute region.}

\item[Figure 7]{ (a) Effect of $\nps$, the size of the PS block, on the 
experimental measured
pressure as function of  $\Sigma/\np$ the
rescaled area/molecule.
(b) Calculation from the SCMF theory for  $\nps= 13$ (smooth line),
$\nps=20$ (dotted line), $\nps=30$ 
(dashed line) and $\nps=50$ (long dashed line). In all 
calculated isotherms 
$\np=100$.  }

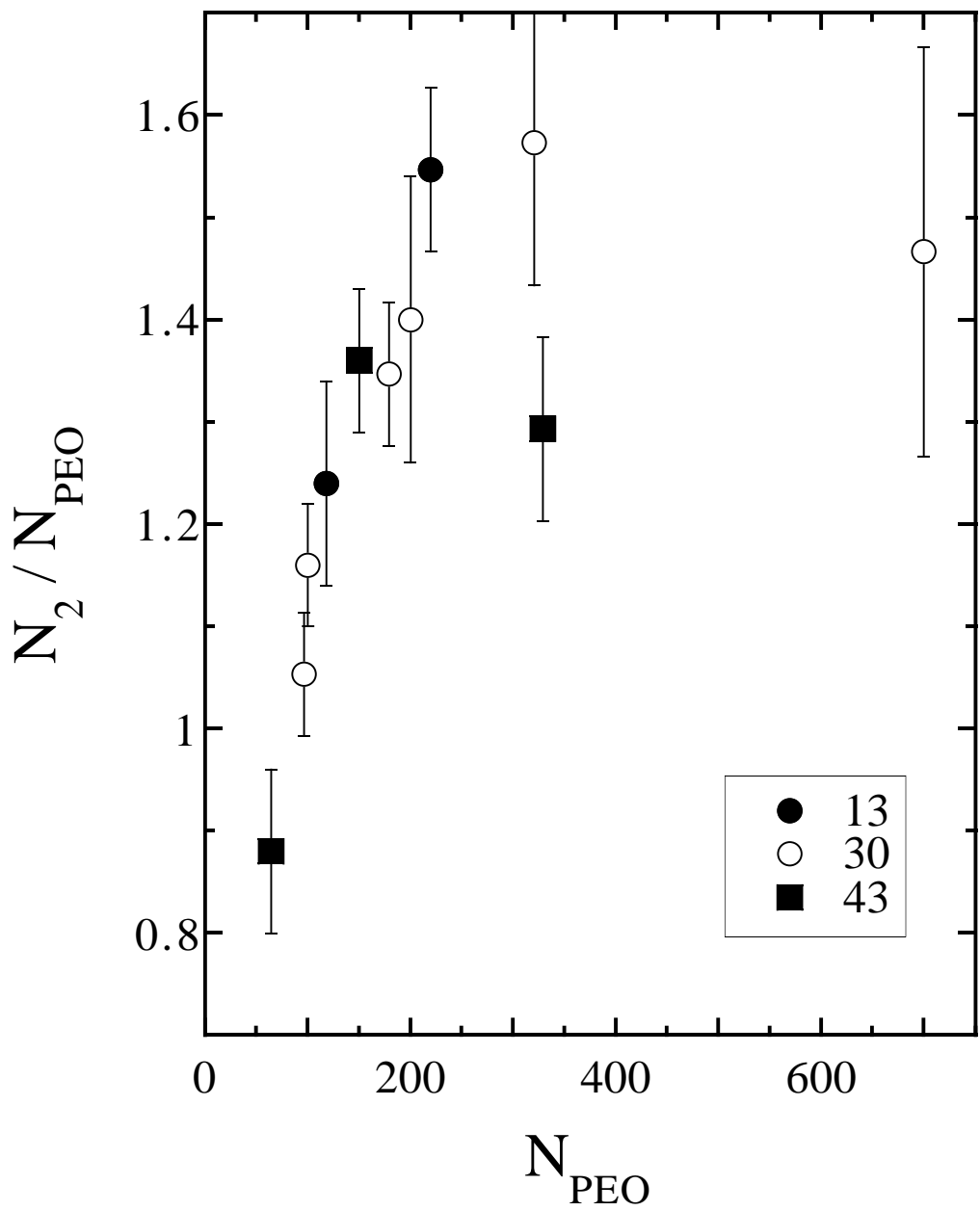
\item[Figure 8]{Fraction of PEO monomers at the interface,
$\npt/\np$, as a function of $\np$, the total PEO chain length.
The value for large $\np$ is larger than one, probably because
of the approximations in the prefactor in (\ref{d9}).}

\end{description}

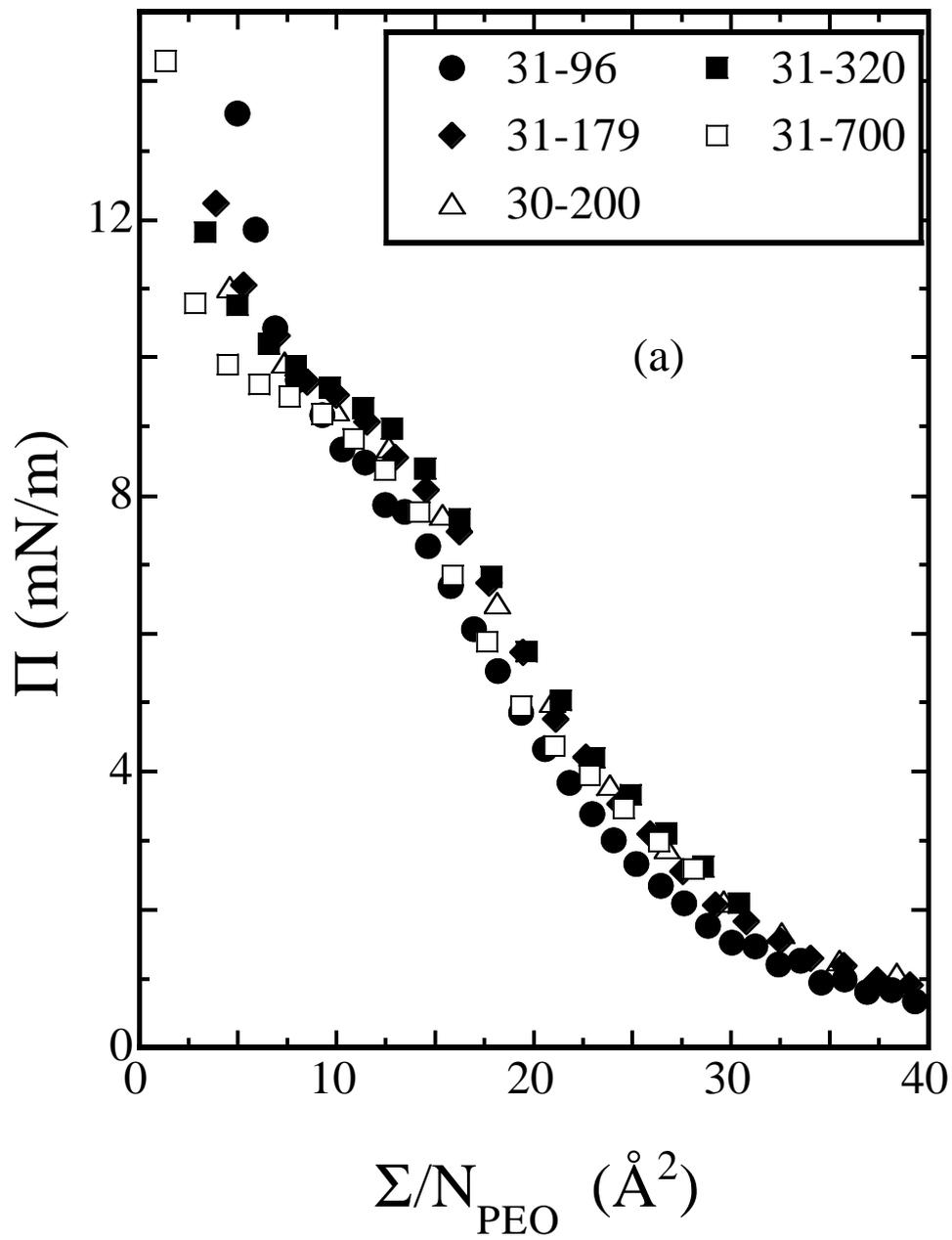
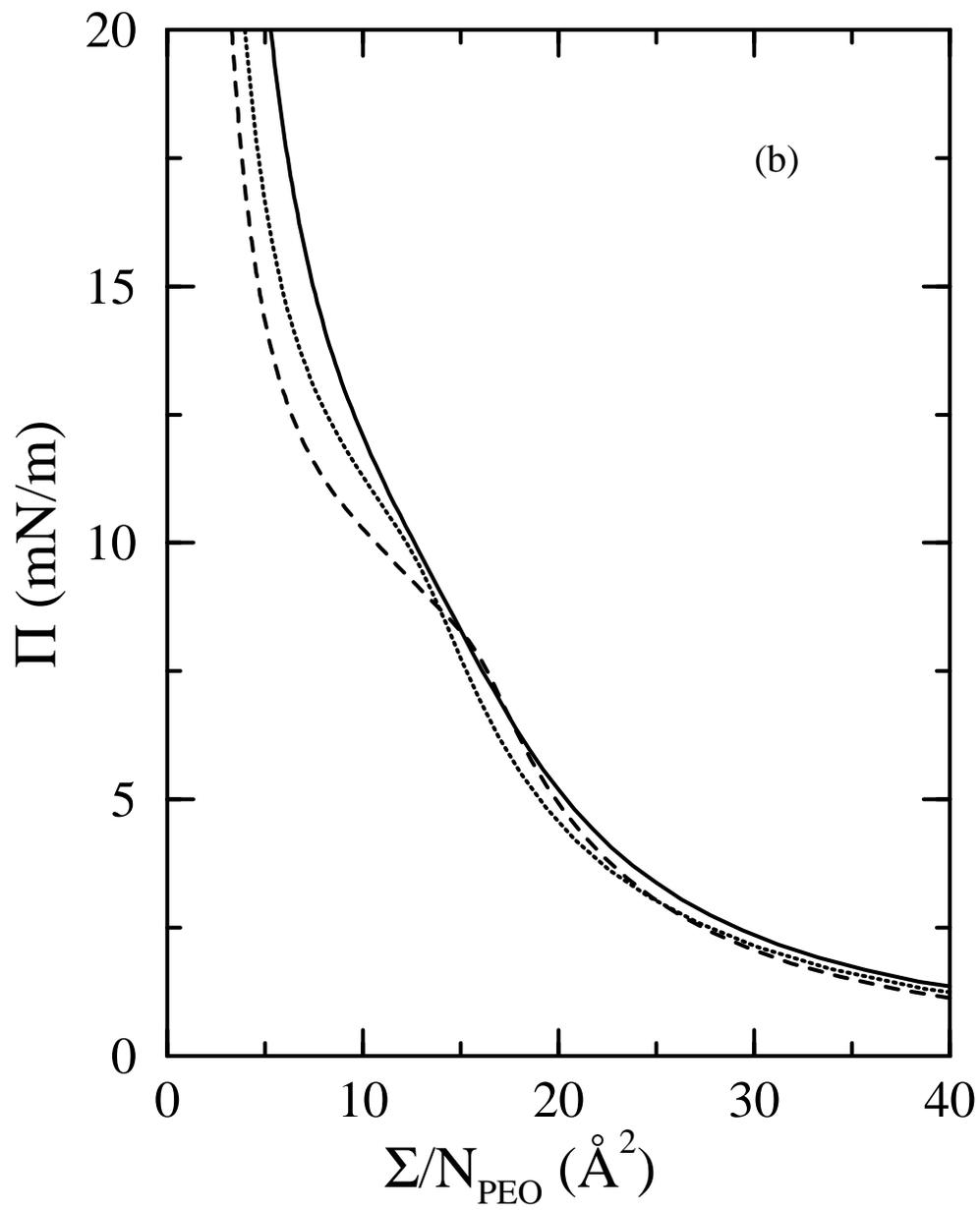
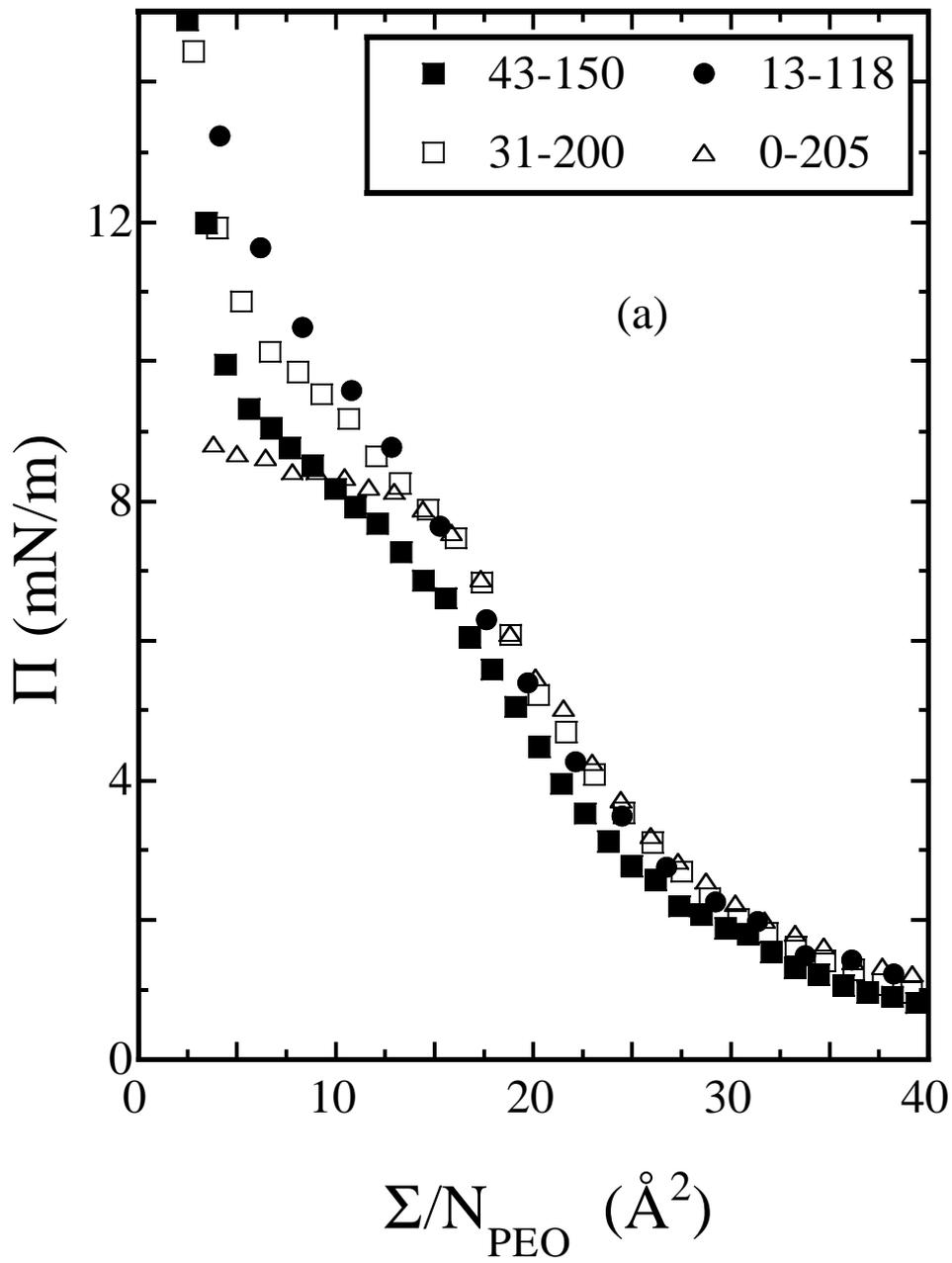
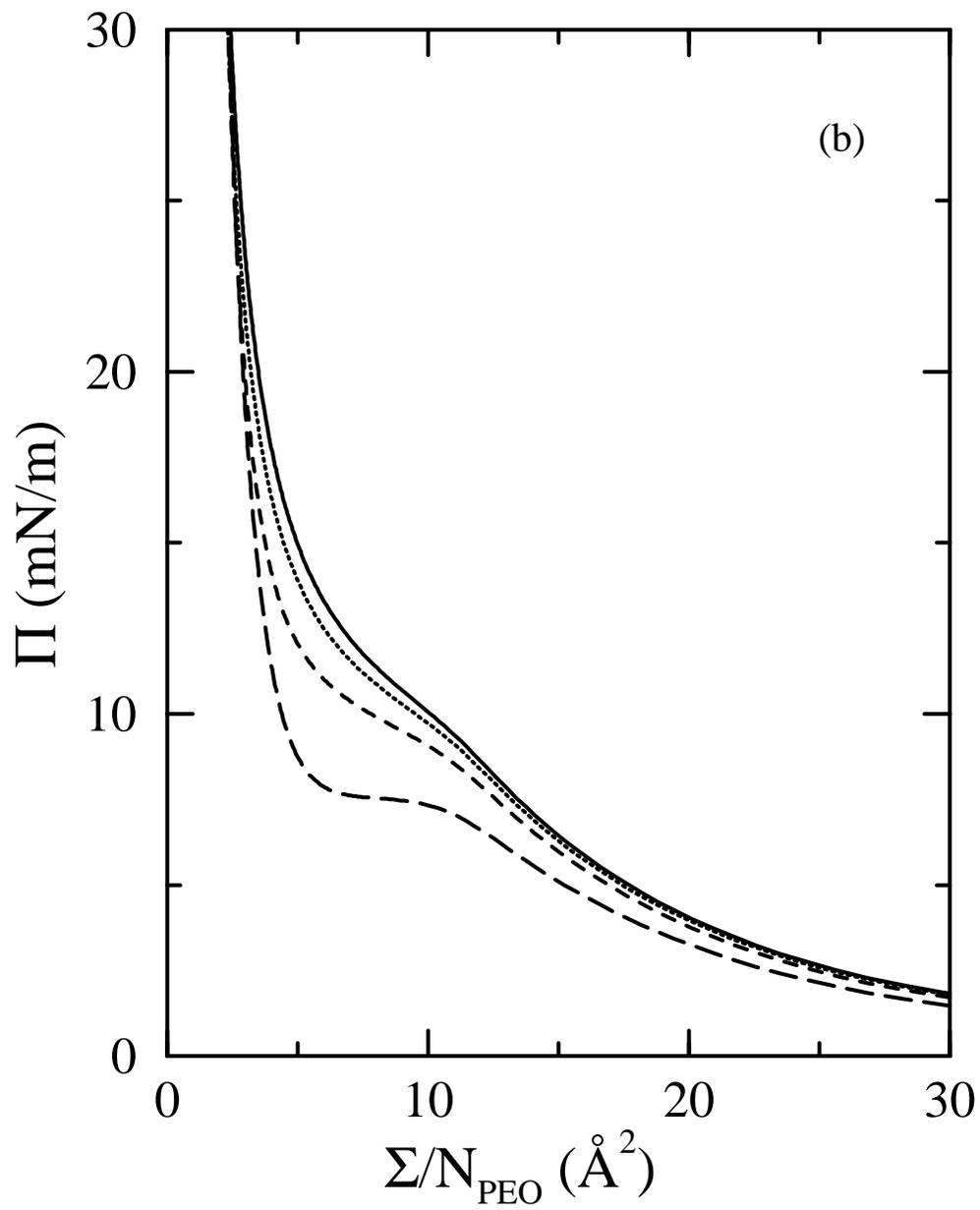
\end{document}